\documentclass[times, twocolumn]{aastex63}
\usepackage{graphicx, graphics}
\usepackage{bbm}
\definecolor{cyan (process) }{rgb}{0.0, 0.72, 0.92}
\usepackage{amsmath, amssymb, bm}
\usepackage[nodayofweek]{datetime}
\usepackage{multirow}

\newcommand{\uat}[2]{\href{http://vocabs.ands.org.au/repository/api/lda/aas/the-unified-astronomy-thesaurus/current/resource.html?uri=http://astrothesaurus.org/uat/#1}{#2 (#1)}}

\shorttitle{Trends in \emph{Spitzer} Secondary Eclipses}
\shortauthors{Wallack, Knutson, \& Deming}

\begin{document}
\pagenumbering{arabic}
\title{Trends in \emph{Spitzer} Secondary Eclipses}
\author{Nicole L. Wallack}
\author{Heather A. Knutson}
\affiliation{Division of Geological \& Planetary Sciences, California Institute of Technology, Pasadena, CA 91125, USA; \url{nwallack@caltech.edu}}

\author{Drake Deming}
\affiliation{Department of Astronomy, University of Maryland at College Park, College Park, MD 20742, USA}

\begin{abstract}
It is well-established that the magnitude of the incident stellar flux is the single most important factor in determining the day-night temperature gradients and atmospheric chemistries of short-period gas giant planets. However it is likely that other factors, such as planet-to-planet variations in atmospheric metallicity, C/O ratio, and cloud properties, also contribute to the observed diversity of infrared spectra for this population of planets. In this study we present new 3.6 and 4.5 micron secondary eclipse measurements for five transiting gas giant planets: HAT-P-5b, HAT-P-38b, WASP-7b, WASP-72b, and WASP-127b. We detect eclipses in at least one bandpass for all five planets and confirm circular orbits for all planets except for WASP-7b, which shows evidence for a non-zero eccentricity. Building on the work of \cite{Garhart2020}, we place these new planets into a broader context by comparing them with the sample of all planets with measured \emph{Spitzer} secondary eclipses. We find that incident flux is the single most important factor for determining the atmospheric chemistry and circulation patterns of short-period gas giant planets. Although we might also expect surface gravity and host star metallicity to play a secondary role, we find no evidence for correlations with either of these two variables.
\end{abstract}


\section{Introduction}
For short-period gas giant planets with hydrogen-rich envelopes, the amount of incident flux received from the star is predicted to be the primary factor that determines the shape of their observed dayside emission spectra (e.g., \citealt{Fortney2008}, \citealt{Burrows2008}). We expect these short-period planets to be tidally locked, where the efficiency of day-night circulation varies as a function of the incident flux (e.g., \citealt{Komacek2016}). This atmospheric circulation in turn determines the temperature of the upper region of the dayside atmosphere, which sets the equilibrium chemistry and corresponding atmospheric composition (e.g., \citealt{heng2018}). Although photochemistry and mixing from the nightside and deep interior can alter this default chemistry, these effects are predicted to be below the sensitivity of current observations for the majority of planets observed to date (e.g., \citealt{Moses2014}). Condensate clouds can also alter the observed atmospheric properties of these planets, but the effects of these clouds are expected to be less pronounced for dayside thermal emission spectra than for transmission spectroscopy (e.g., \citealt{Fortney2005}).

Observations of the secondary eclipse, when the planet passes behind its host star, allow us to probe the thermal emission spectra of transiting gas giant planets. This has enabled detailed studies of a handful of planets (e.g., \citealt{Brogi2017}, \citealt{Morley2017}, \citealt{Kreidberg2018}), but there are relatively few planets with such extensive secondary eclipse data sets. If we broaden our focus to planets with just a few broadband photometric measurements from \emph{Spitzer}, we can search for broader population-level trends. Previous studies of dayside emission spectra confirm that hotter planets do indeed have less efficient heat redistribution (\citealt{Schwartz2015}, \citealt{Garhart2020}), in good agreement with predictions from atmospheric circulation models (\citealt{Perez-Becker2013}, \citealt{Komacek2016}). \cite{Garhart2020} additionally found evidence for a systematic shift in the $3.6$ to $4.5$~$\mu$m spectral slopes of these planets as a function of incident flux, which suggests that the atmospheric chemistries and pressure-temperature profiles of these planets also vary as a function of the irradiation. However, this study found that neither of the two most commonly utilized model atmosphere grids was able to accurately predict the increase in the observed ratio of 3.6 $\mu$m and 4.5 $\mu$m brightness temperatures with increasing equilibrium temperature, suggesting that these models can be further improved. 

To date, most published population-level studies of transiting gas giant planet emission spectra have focused on searching for correlations with the incident flux (e.g., \citealt{Cowan2011}, \citealt{Schwartz2015}, \citealt{Schwartzetal.2017}, \citealt{Garhart2020}, \citealt{Baxter2020}). However, we expect that planets with the same incident flux levels might nonetheless possess distinct thermal spectra if they have different atmospheric metallicities and/or surface gravities, both of which can alter their atmospheric chemistries, circulation patterns, and cloud properties. In a previous study \citep{Wallack2019} we focused on the sub-population of planets cooler than $\sim1000$~K, which are expected to undergo a particularly distinct shift in atmospheric chemistry as a function of atmospheric metallicity and C/O ratio (e.g., \citealt{Moses2013}, \citealt{Drummond2018}). In this study, we broaden our focus to the full sample of transiting gas giant planets with \emph{Spitzer} secondary eclipse detections in order to determine whether or not there are additional parameters beyond incident flux that might help to explain the observed diversity of dayside emission spectra. These same factors might also provide new insights into how to modify standard atmosphere model grids in order to better match the observed trends in spectral shape as a function of incident flux.

\begin{centering}
\begin{deluxetable*}{lccccc}[!t]
\tablecaption{System Properties for New Planets in this Study}
\tablewidth{0pt}
\tablehead{
 \colhead{}&\colhead{HAT-P-5b}&\colhead{HAT-P-38b} &\colhead{WASP-7b}&\colhead{WASP-72b}&\colhead{WASP-127b}}
\startdata
T$_*$ (K) & 5960$\pm$100&5330$\pm$100&6520$\pm$70&6250$\pm$100&5750$\pm$100\\
$[$Fe/H$]_{*}$& 0.24$\pm$0.15&0.06$\pm$0.1&0.00$\pm$0.10&-0.06$\pm$0.09&-0.18$\pm$0.06\\
Mass (M$_{\tt Jup}$) & 1.06 $\pm$0.11&0.267$\pm$0.020&0.98$\pm$0.13&1.461$^{+0.059}_{-0.056}$&0.165$^{+0.021}_{-0.017}$\\
Radius (R$_{\tt Jup}$) & 1.252$\pm$0.043&0.825$^{+0.092}_{-0.063}$&1.374$\pm$0.094&1.27$\pm$0.20&1.311$^{+0.025}_{-0.029}$\\
T$_{\tt equ}$ (K) \tablenotemark{a}& 1517$\pm$37&1080$^{+60}_{-45}$&1530$\pm$50&2204$^{+139}_{-115}$&1404$\pm$29\\
$e$\tablenotemark{b}&$<$0.072 ($<$0.18) &$<$0.055 ($<$0.17) &$<$0.049 ($<$0.11) &$<$0.017 ($<$0.038) &0\\
$\omega$ (deg) \tablenotemark{b}&...&...&...&...&...\\
Period (days) \tablenotemark{c}&2.78847360 (52) &4.640382 (32) & 4.9546416 (35) &2.2167421 (81) &4.17807015 (57) \\
T$_{c}$ (BJD-2,450,000) \tablenotemark{c} &5432.45510 (10) &5863.12034 (35) &5446.63493 (30) & 5583.6528 (21) &7248.741276 (68) \\
References&1,2,3,4,5&1,5,6&1,5,7,8&1,5,9&5,10,11
\enddata 
\label{table:systems1}
\tablecomments{
\tablenotetext{a}{Calculated assuming planet-wide heat circulation and zero albedo.}
\tablenotetext{b}{The orbital eccentricity $e$ and longitude of periapse $\omega$ are derived from fits to radial velocity data.} 
\tablenotetext{c}{Uncertainties on the last two digits are parenthesized.} 
}
\tablerefs{
 (1) \citet{Bonomo2017a}, (2) \citet{Southworth2012}, (3) \citet{Bakos2007}, (4) \citet{Torres2008}, (5) \citet{Southworth2011}, (6) \citet{sato2012}, (7) \citet{Albrecht2012}, (8) \citet{Hellier2009}, (9) \citet{Gillon2013a}, (10) \citet{Lam2017}, (11) \citet{chen2018}
}
\end{deluxetable*}
\end{centering}

\cite{Garhart2020} presented a uniform analysis of 3.6 and 4.5 $\mu$m \emph{Spitzer} observations of 36 transiting hot Jupiters. We use 31 of these observations (those with detections above 2.5$\sigma$) and expand on this sample by leveraging an additional 42 planets with \emph{Spitzer} 3.6 and 4.5 $\mu$m secondary eclipse detections (above 2.5$\sigma$) from the literature as well as adding secondary eclipse measurements for three of our five new planets: HAT-P-5b \citep{Bakos2007}, WASP-7b \citep{Hellier2009}, and WASP-127b \citep{Lam2017} (HAT-P-38b \citep{sato2012} and WASP-72b \citep{Gillon2013a} do not have detections in both bandpasses). With this newly expanded sample we proceed to revisit previously established correlations between spectral shape and incident flux, and to search for additional correlations with stellar metallicity and surface gravity. In Section \ref{sec:observations} we describe our photometric extraction and model fits. In Section \ref{sec:results} we present new \emph{Spitzer} secondary eclipse measurements of five new planets, and in Section \ref{sec:discussion}, we add these new planets to the published \emph{Spitzer} secondary eclipse measurements and investigate trends in the thermal emission spectra of the population of short-period gas giant planets.

\section{Observations and Data Analysis}\label{sec:observations}

We obtained one secondary eclipse each in the IRAC 3.6 $\mu$m and 4.5 $\mu$m bands \citep{Fazio2004} for HAT-P-5b (PID: 60021), WASP-7b (PID: 60021), WASP-72b (PID: 10102), and WASP-127b (PID: 13044) and two visits in each band for HAT-P-38b (PID: 12085). Aside from HAT-P-5b, all of these data were taken in the $32\times32$ pixel subarray mode with an initial 30 minute observation to allow for settling of the telescope followed by a peak-up pointing adjustment prior to the start of the science observation \citep{Ingalls2012}. HAT-P-5b and WASP-7b are the oldest data sets in this study and were observed before the peak-up pointing mode was fully implemented. Additionally, HAT-P-38b was observed in full array mode. See Table~\ref{table:systems1} and Table~\ref{table:observations} for additional details.

\begin{centering}
\begin{deluxetable*}{ccccccccccc}[t!]
\tablecaption{\emph{Spitzer} Observation Details}
\tablewidth{0pt}
\tablehead{
\colhead{Target} & \colhead{$\lambda$ ($\mu$m) } & \colhead{UT Start Date}&{AOR}& \colhead{Length (h) } & \colhead{t\textsubscript{int} (s) \tablenotemark{a}} &
\colhead{t\textsubscript{trim} (h) \tablenotemark{b}} & \colhead{r\textsubscript{pos}\tablenotemark{c} } & \colhead{r\textsubscript{phot}\tablenotemark{d}} & \colhead{n$\textsubscript{bin}$\tablenotemark{e}} & RMS\tablenotemark{f}}
\startdata
HAT-P-5b &  3.6 & 2009 Oct 16&31757056&7.64&6.0&0.5&3.0&2.0&2&1.25\\
        &  4.5 & 2009 Oct 19&31751424&7.64 &6.0&2.0&3.0&2.0&4&1.26\\
HAT-P-38b &  3.6 & 2016 Apr 10 &58238976&8.91&2.0&1.0&4.0&2.0&4&1.15\\
        &  3.6 & 2016 May 08 &58241024&8.95& 2.0 & 1.5&4.0&2.0&2&1.14\\
        &  4.5 & 2016 Apr 20 &58238464&8.95& 2.0 & 2.0&3.5&2.0&2&1.17\\
        &  4.5 & 2016 May 04 &58241280&8.95& 2.0 & 2.0&3.5&2.0&32&1.20\\
WASP-7b &  3.6 & 2010 Jun 12 &31770880&7.73 & 2.0 &0.5\tablenotemark{g}&3.0&5.0&2&1.39\\
         &   4.5 &2010 Jun 27&31765248&7.73&2.0 &0.5\tablenotemark{g}&3.0&2.9&16&1.22\\
WASP-72b &  3.6 & 2014 Nov 17&51816192& 10.11&0.4 &1.0&4.0&2.0&512&1.31\\
         &   4.5 &2014 Nov 19&51842304&10.11 &0.4 & 1.0&3.5&2.0&64&1.53\\
WASP-127b &  3.6 & 2017 Aug 20&62161664& 12.55& 2.0 & 2.0 & 4.0 & 2.0& 2& 1.27 \\
          &   4.5 &2017 Sep 01&62162176&12.55 &2.0 &2.0 &2.5 &2.7 &16 & 1.20
\enddata
\tablecomments{
\tablenotetext{a}{Integration time}
\tablenotetext{b}{Initial trim duration }
\tablenotetext{c}{Radius of the aperture (in pixels) used to determine the location of the star on the array}
\tablenotetext{d}{Radius of the aperture (in pixels) used for the photometry}
\tablenotetext{e}{Bin size used for fits}
\tablenotetext{f}{Ratio of measured RMS to photon noise limit}
\tablenotetext{g}{WASP-7b has a time of secondary eclipse that is not well centered in the observation window, therefore in order to preserve as much of ingress as possible, we fix the trim duration to 30 minutes (see \S\ref{sec:observations} for more details).}
}
\label{table:observations}
\end{deluxetable*}
\end{centering}

We utilize the standard Basic Calibrated Data (BCD) images for our analysis and extract photometric fluxes as described in our previous studies (i.e., \citealt{Wallack2019}). In brief, we first calculate the BJD\textsubscript{UTC} mid-exposure times for each image, then estimate the sky background in each image by masking out a circular region with a radius of 15 pixels centered on the position of the star, iteratively trimming $3\sigma$ outliers, and fitting a Gaussian function to a histogram of the remainder of the pixels. We utilize flux-weighted centroiding (e.g., \citealt{Knutson2008}; \citealt{Deming2015}) with a circular aperture to determine the location of the star on the array, considering aperture radii ranging between $2.5-4.0$ pixels in 0.5 pixel steps and optimize our choice of aperture as described below. We use the \texttt{aper} routine in the \texttt{DAOPhot} package \citep{Stetson1987DAOPHOT:APhotometry} to extract the total flux in a circular aperture centered on the position of the star, considering aperture sizes ranging from $2.0-3.0$ pixels in steps of 0.1 pixels and from $3.0-5.0$ pixels in steps of 0.5 pixels.

In order to mitigate the ramp-like behavior present at early times in some of the visits, we trim up to two hours of data from the beginning of each time series. We find that binning our data prior to fitting reduces the amount of time-correlated noise in the residuals (see \citealt{Deming2015} and \citealt{Kammer2015} for more details). In order to determine the optimal combination of flux-weighted centroiding aperture, photometric aperture, trim duration, and bin size for each visit, we fit a combined instrumental and astrophysical model to each version of the photometry and calculate the standard deviation of the residuals as a function of bin size stepping in powers of two (see \citealt{Kammer2015} for further details). We then calculate the least-squares difference between the measured standard deviation of the residuals and the predicted photon noise limit in each bin, which decreases as the square root of the number of points. We then select the photometric and centroiding apertures, trim duration, and bin size that minimizes this least-squares difference (i.e., the one that is closest to the photon noise at all measured timescales) for use in our subsequent analysis. 

Our model for each visit consists of a secondary eclipse light curve and an instrumental noise model, which we fit simultaneously. We calculate our eclipse model using the \texttt{batman} package (\citealt{Mandel2002}, \citealt{Kreidberg2015}), 
where we fix the planet-star radius ratio, orbital inclination, and the ratio of the orbital semi-major axis to the stellar radius ($a/R_*$) to the published values for each planet (see Table ~\ref{table:systems1} for references) and allow the eclipse depth and time to vary as free parameters. Due to the fact that the orbital parameters are often more precisely measured from transit light curves than from secondary eclipse light curves, we are justified in fixing the orbital parameters to those measured from transit light curves instead of letting these parameters vary in our fits.

The dominant instrumental noise source for \emph{Spitzer} timeseries photometry is intra-pixel sensitivity variations (\citealt{Reach2005}, \citealt{Charbonneau2005}, \citealt{Morales-Calderon2006}), which cause the apparent flux from the star to vary as a function of its position on the pixel. We model this effect using the pixel-level decorrelation (PLD) method, which uses a linear combination of individual pixel-level light curves to account for trends due to variations in both the star's position on the array and the width of the stellar point spread function \citep{Deming2015}. As in \citet{Deming2015}, we utilize a 3 $\times$ 3 grid of pixels centered on the location of the star and remove astrophysical flux variations in each 3 $\times$ 3 postage stamp by dividing by the summed flux across all nine pixels. 

In the majority of our observations, we can fully account for additional time-dependent trends with the inclusion of a linear function of time once we have trimmed some data from the start of the observation. This is not true, however, for the 3.6 $\mu$m observation of WASP-72b or the 3.6 $\mu$m observation of WASP-7b. For the 3.6 $\mu$m observation of WASP-72b, we obtain the best fit using an exponential function of time. Using this exponential reduces the Bayesian Information Criterion (BIC) by 12. For the 3.6 $\mu$m observation of WASP-7b, we obtain the best fit using a quadratic function of time ($\Delta$BIC of 66). For all fits we divide out our initial astrophysical model and use linear regression on the residuals to obtain an initial guess for the nine linear PLD coefficients in order to speed up convergence for these highly correlated parameters.

When optimizing our choice of photometry, we first fix the predicted time of eclipse to an orbital phase of 0.5 and run fits on each version of the photometry using a Levenberg-Marquardt minimization. We then select the optimal version of the photometry in each bandpass and carry out a simultaneous fit to all of the visits for a given planet, where we allow the orbital phase of the secondary eclipse to vary as a free parameter. We carry out these fits using the affine-invariant Markov chain Monte Carlo (MCMC) ensemble sampler \texttt{emcee} (\citealt{Foreman-Mackey2013}, \citeyear{Foreman-Mackey2019}), where we allow the secondary eclipse depth to vary independently in each bandpass but assume a common eclipse phase. We place uniform priors on all free parameters and allow the eclipse depths to take on negative values so that we do not bias our eclipse depth estimates. We utilize 60 walkers for our fits, which is enough to ensure adequate sampling of the model parameter space. We initialize these walkers in a tight cluster centered on the best-fit solution from a joint Levenberg-Marquardt minimization and carry out an initial burn-in with a length of 10,000 steps. We then discard this initial burn-in and carry out a subsequent fit with 10$^5$ steps per chain. We then return to the original set of photometry options and repeat our optimization fixing the time of secondary eclipse to the median value from the MCMC chains. We adopt the resulting optimal photometry choices for each visit and rerun the MCMC for the joint fits. 

\begin{figure*}[htb!]
\begin{centering}
\includegraphics[width=.75\textwidth]{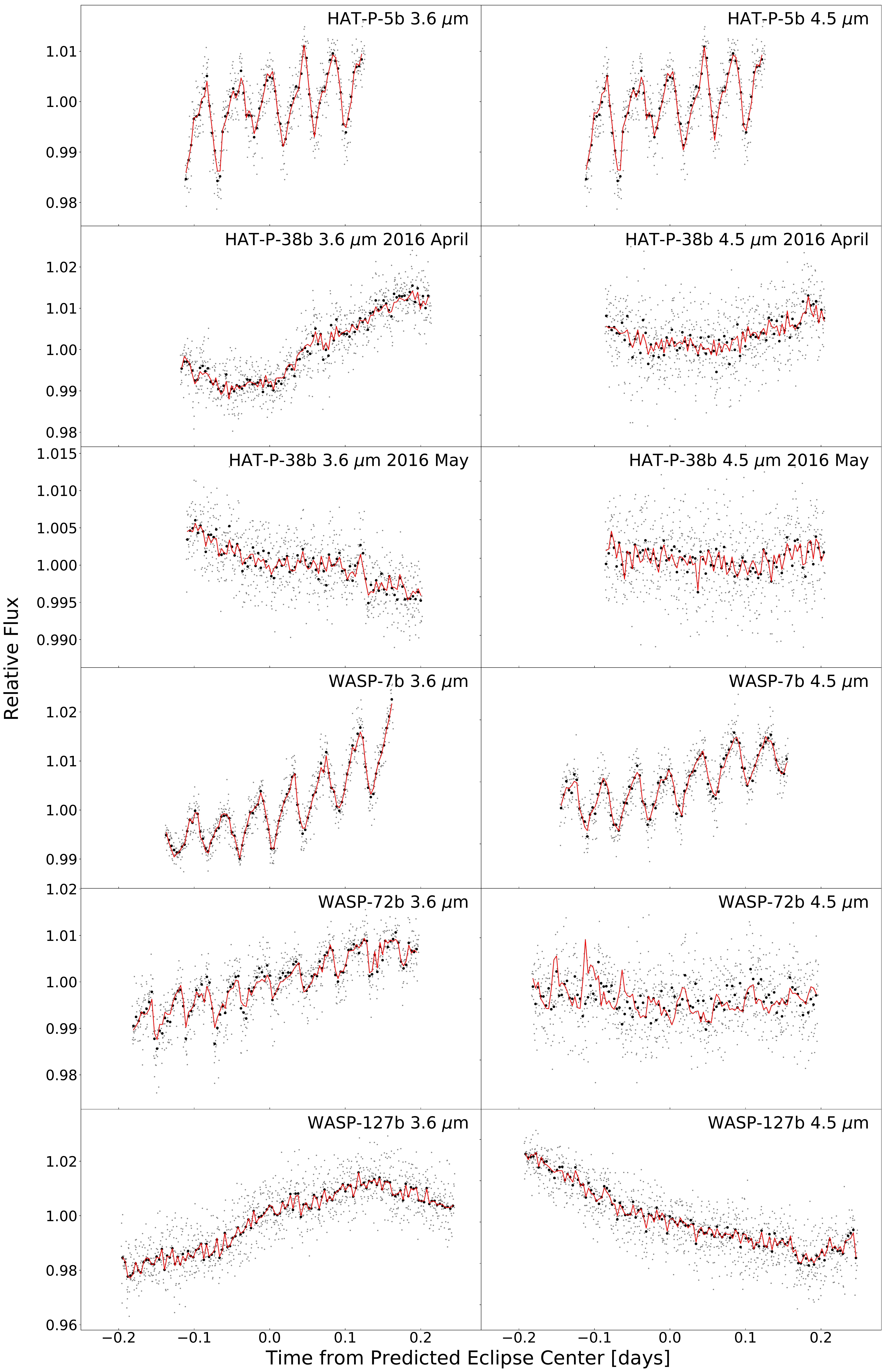}
\caption{Raw \emph{Spitzer} photometry for each visit of HAT-P-5b, HAT-P-38b, WASP-7b, WASP-72b, and WASP-127b. The normalized flux binned in 5 minute intervals is shown as black filled circles and the 30 second binned flux is shown as gray filled circled. The best-fit instrumental model is overplotted in red. Observations of HAT-P-38b are shown chronologically down each column.}
\label{fig:instra} 
\end{centering}
\end{figure*}

\begin{figure*}[htb!]
\begin{centering}
\includegraphics[width=.75\textwidth]{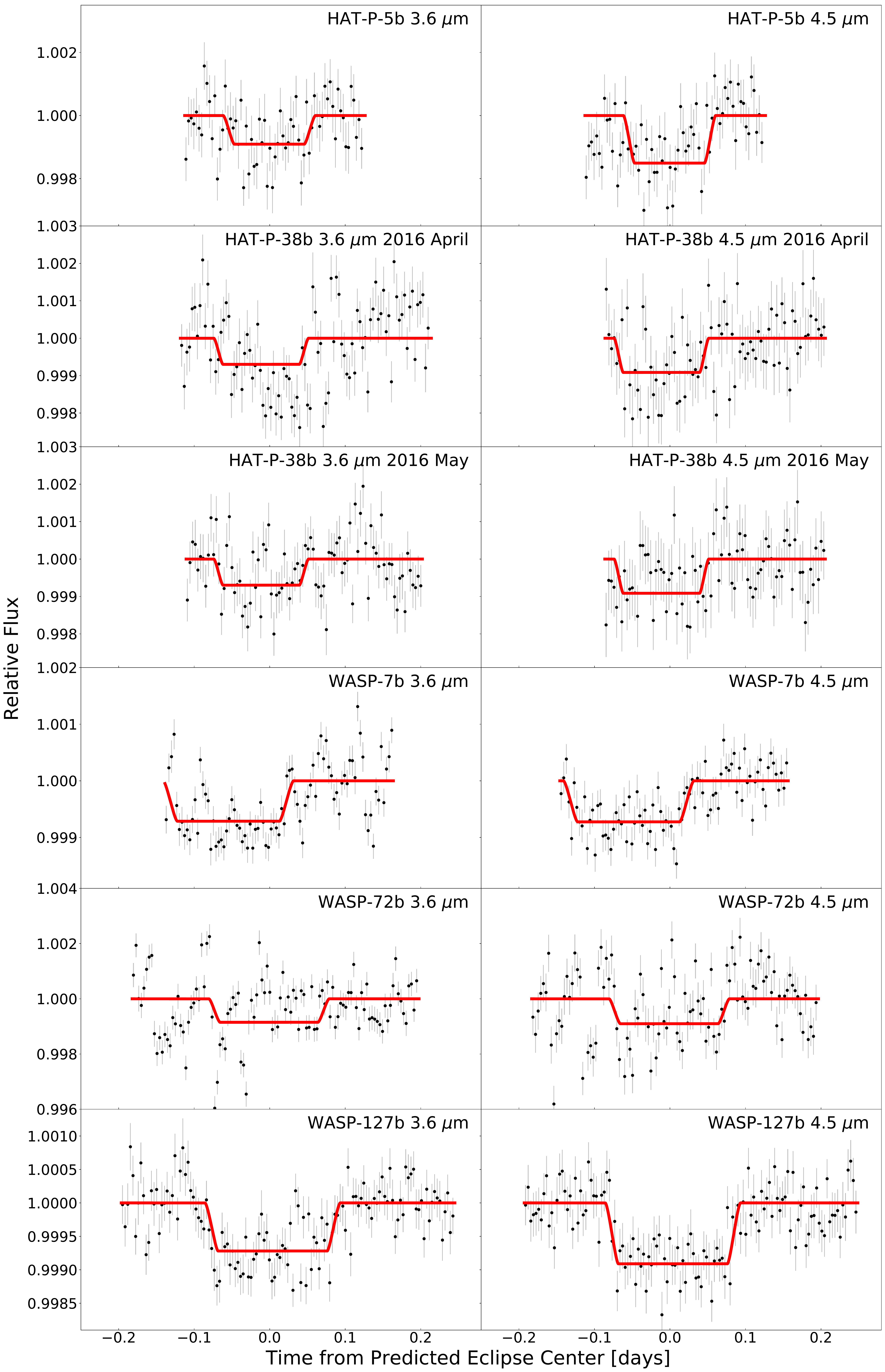}
\caption{Normalized light curves for each visit of HAT-P-5b, HAT-P-38b, WASP-7b, WASP-72b, and WASP-127b from the simultaneous fits with instrumental effects removed. We show data binned in five-minute intervals (black filled circles) with error bars corresponding to the scatter in each bin divided by the square root of the number of points in each bin, and we overplot the best-fit secondary eclipse model in red. Observations of HAT-P-38b are shown chronologically down each column. The 2$\sigma$ upper limits for the best-fit eclipse depths of the 4.5 $\mu$m visits of HAT-P-38b and the 3.6 $\mu$m visit of WASP-72b are shown.}
\label{fig:lc}   
\end{centering}
\end{figure*}

\begin{figure*}[htb!]
\begin{centering}
\includegraphics[width=.75\textwidth]{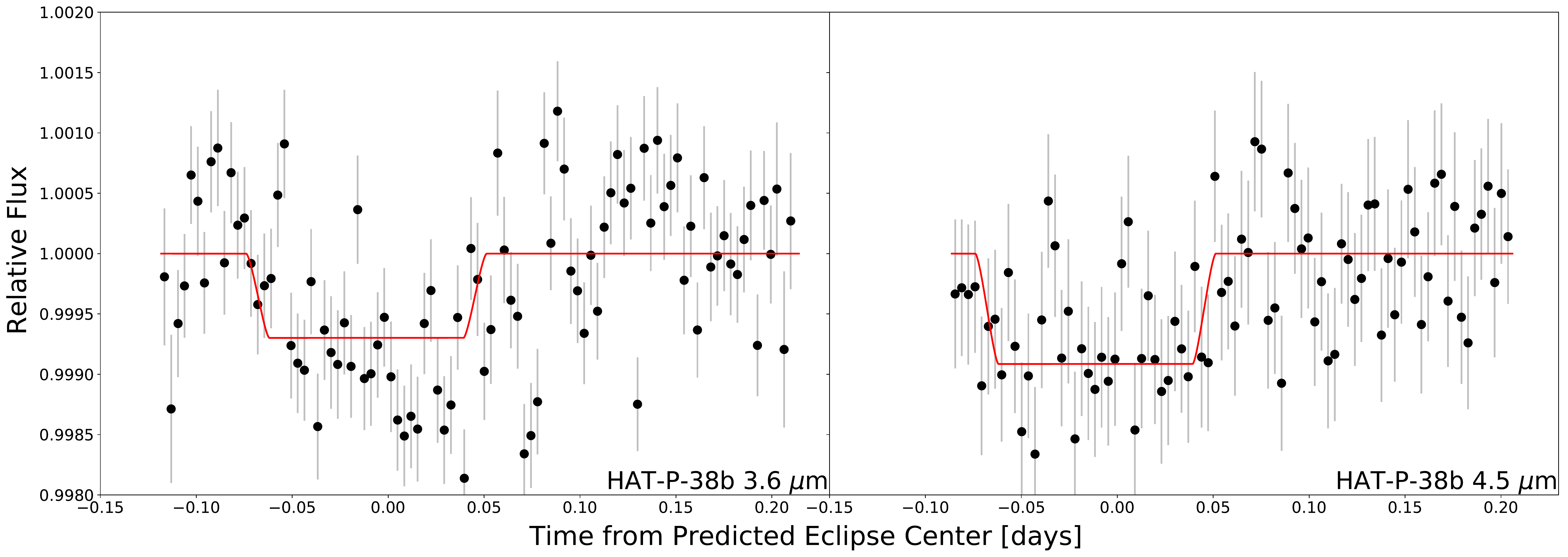}
\caption{Band-averaged light curves for HAT-P-38b from the simultaneous fits with instrumental effects removed. We show data binned in five-minute intervals (black filled circles) with error bars corresponding to the scatter in each bin divided by the square root of the number of points in each bin, and overplot the best-fit eclipse model in each bandpass for comparison (red lines). The 2$\sigma$ upper limit for the best-fit eclipse depth of the 4.5 $\mu$m data is shown.}
\label{fig:lc_h38}   
\end{centering}
\end{figure*}

\begin{figure*}[htb!]
\begin{centering}
\includegraphics[width=.75\textwidth]{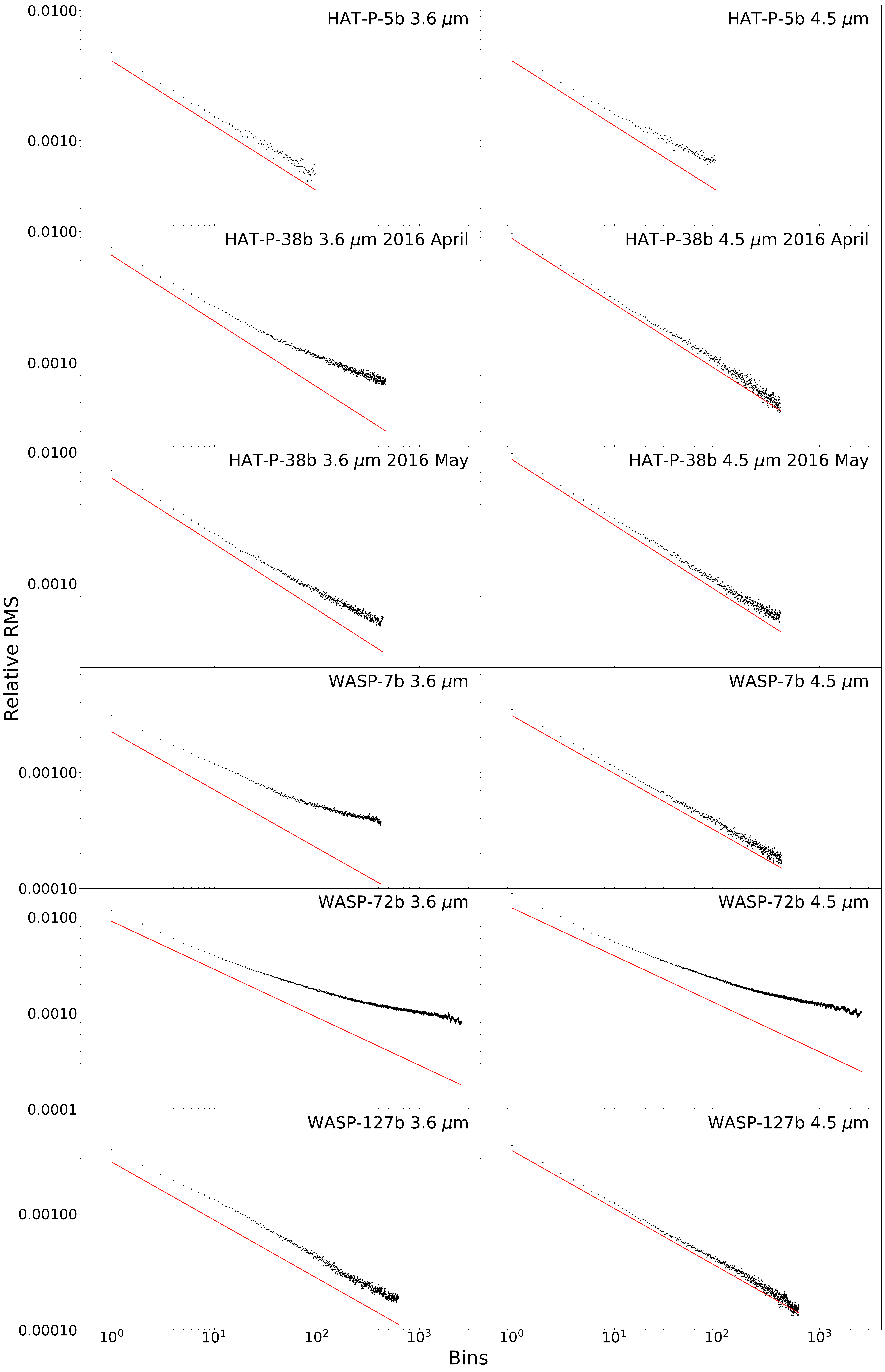}
\caption{Standard deviation of the residuals as a function of bin size after removing the best-fit instrumental and astrophysical models. The solid lines show the predicted photon noise limit as a function of bin size, which follows a 1/$\sqrt{N}$ scaling. Observations of HAT-P-38b are shown chronologically down each column.}
\label{fig:rms}   
\end{centering}
\end{figure*}

We report the median values from our MCMC chains and the corresponding 1$\sigma$ uncertainties in Table~\ref{table:bestfit} and show the raw photometry for each visit with best-fit instrumental noise models from the joint fits overplotted in Figure~\ref{fig:instra}. Normalized light curves for these visits with best-fit eclipse light curves overplotted are shown in Figure~\ref{fig:lc}. In Figure~\ref{fig:lc_h38} we combine all visits for HAT-P-38b and show the averaged light curves for each bandpass. The standard deviation of the residuals as a function of bin size for all visits are shown in Figure~\ref{fig:rms}.

We alter our fitting procedure for WASP-7b, as there appears to be substantial correlated noise (i.e. the residuals do not scale with $\sqrt{n}$) in the residuals of the 3.6 $\mu$m data (see the WASP-7b 3.6 $\mu$m panel in Figure~\ref{fig:rms}). To mitigate any biases in our best-fit parameters, we initially fit each of the channels for WASP-7b independently. We find that the best-fit secondary eclipse phases for each channel are consistent at the 1 sigma level, indicating that the correlated noise in the 3.6 $\mu$m data is likely not biasing our time of secondary eclipse in that channel. 

We find that both the 3.6 and 4.5 $\mu$m data prefer an eclipse phase that is offset from the expected value for a circular orbit (see Section~\ref{sec:results} for more details). As a result, the secondary eclipse is not centered in the observation, but instead occurs 78.8$^{+5.0}_{-4.2}$ minutes early. In order to preserve as much of ingress as possible, we only trim 30 minutes from the beginning of each observation for this planet rather than considering a range of trim durations and optimizing to minimize the scatter in the residuals. We account for the effect of correlated noise on the 3.6 $\mu$m eclipse depth uncertainty by inflating the per-point errors (which are generally left as a free parameter in our fits) by a factor that reflects how much the variance of the residuals deviates from the expected white noise scaling (i.e. $1/\sqrt{n}$ where n is the number of points in each bin) at a characteristic timescale of 10 minutes (see \citealt{Pont2006} and \citealt{Lanotte2014} for more information). Because we are fitting binned light curves, we calculate this inflation factor as the amount of excess noise relative to the expected $\sqrt{n}$ scaling when we go from the binning timescale used in the fits to a binning timescale of 10 minutes (see Figure~\ref{fig:rms}). 
We find that the resulting inflation factor is 2.1 for the 3.6 $\mu$m WASP-7b observation. We then take our best fit per-point error from the initial fit, multiply it by that factor, fix the per-point uncertainty to that value, and rerun our fit in order to obtain an updated eclipse depth and phase. 

It is apparent in Figure~\ref{fig:rms} that several other observations also appear to have excess correlated noise. We calculate the inflation factor for each observation following the same process as described above, and implement a new version of the fit with an inflated per-point uncertainty for visits with inflation factors larger than 1.5. We find that correlated noise exceeding this threshold is present in both the 3.6 and 4.5 $\mu$m observations of WASP-72b (inflation factors of 1.5 and 2.0, respectively; see Figure~\ref{fig:rms}) and both 3.6 $\mu$m observations of HAT-P-38b (inflation factors of 1.7 and 1.5 for the first and second observations respectively). For WASP-72b, we do not detect the eclipse in the 3.6 $\mu$m bandpass, so we cannot compare the best-fit secondary eclipse phases from each channel in order to determine if this parameter is affected by correlated noise. However, the best-fit secondary eclipse phase from the 4.5 $\mu$m fit agrees with the prediction for a circular orbit, and the alternative scenario (slightly eccentric orbit biased by correlated noise to appear circular) seems unlikely.

\section{Results}\label{sec:results}

\begin{centering}
\begin{deluxetable*}{ccccccc}[t!]
\tablecaption{Best Fit Eclipse Parameters}
\tablewidth{0pt}
\def\arraystretch{1.0}
\tablehead{
\colhead{Target} & \colhead{Band ($\mu$m) } & \colhead{Depth (ppm) } &\colhead{T\textsubscript{Bright} (K) } & \colhead{ Time Offset (days) \tablenotemark{a}} & Center of Eclipse (Phase) & \colhead{ecos ($\omega$) \tablenotemark{b}}}
\startdata
HAT-P-5b    &  3.6 &908$^{+202}_{-201}$&1485$^{+109}_{-118}$& -0.0006 $^{+0.0024}_{-0.0022}$& 0.4998$^{+0.0009}_{-0.0008}$&-0.0003$^{+0.0014}_{-0.0012}$\\
              &  4.5 & 1508$\pm$266 &1567$^{+115}_{-121}$\\
HAT-P-38b    &  3.6 & 698$\pm$189\tablenotemark{c}  &1503$^{+135}_{-150}$&-0.0113$^{+0.0067}_{-0.0042}$\tablenotemark{e} &0.4976$^{+0.0014}_{-0.0009}$ & -0.0038$^{+0.0023}_{-0.0014}$  \\
 &  4.5 & $<$914\tablenotemark{d}& $<$1425\\
WASP-7b  &  3.6 & 714$^{+191}_{-190}$\tablenotemark{c}&1583$^{+147}_{-161}$& -0.0547$^{+0.0035}_{-0.0029}$ &0.4890$^{+0.0007}_{-0.0006}$& -0.0173$^{+0.0011}_{-0.0009}$ \\
&   4.5 & 725$^{+109}_{-106}$&1393$^{+80}_{-82}$\\
WASP-72b  &  3.6 &$<$852$\tablenotemark{c}$$^{,}$$\tablenotemark{d}$&$<$2265&-0.0009$^{+0.0054}_{-0.0071}$&0.4996$^{+0.0024}_{-0.0032}$&  -0.0006$^{+0.0038}_{-0.0050}$ \\
&  4.5 &  903$^{+288}_{-294}$\tablenotemark{c}&$2098^{+335}_{-364}$\\
WASP-127b  &  3.6 & 719$\pm$62& 1454$_{-43}^{+42}$&0.0038$^{+0.0013}_{-0.0015}$&0.5009$^{+0.0003}_{-0.0004}$  &  0.0014$^{+0.0005}_{-0.0006}$  \\
 &   4.5 & 910 $\pm$69 &1373$_{-41}^{+40}$
\enddata 
\tablecomments{
\tablenotetext{a}{Time offset from predicted center of eclipse. We fit both channels with a common time of secondary eclipse.}
{\tablenotetext{b}{Computed using the approximation for a low eccentricity orbit.}}
\tablenotetext{c}{There is correlated noise present in the residuals of this fit (see Figure~\ref{fig:rms}), and we therefore inflate the per-point errors in our fits (see \S\ref{sec:observations} for more information).} 
\tablenotetext{d}{We report the 2$\sigma$ upper limit for the eclipse depth.}
\tablenotetext{e}{HAT-P-38b has a bimodal distribution for the time of secondary eclipse with one peak consistent with a circular orbit and a second smaller peak offset by $\sim$30 minutes. We report the eclipse depths and time for the solution that is consistent with a circular orbit; see \S\ref{sec:results} for more details.}
}
\label{table:bestfit}
\end{deluxetable*}
\end{centering}
We report the best-fit eclipse depths and times and their corresponding uncertainties in Table~\ref{table:bestfit}. We detect the eclipse in both bandpasses with greater than $3\sigma$ significance for HAT-P-5b, WASP-7b, and WASP-127b. For HAT-P-38b we detect the eclipse at 3.6 $\mu$m but not at 4.5 $\mu$m, and for WASP-72b we detect the eclipse at 4.5 $\mu$m but not at 3.6 $\mu$m. This allows us to place relatively tight constraints on the eclipse depth in the bandpass with the non-detection, as the eclipse phase is effectively fixed in the joint fit by the detection in the other bandpass. We find that the best-fit eclipse phases for HAT-P-5b, HAT-P-38b, WASP-72b, and WASP-127b are all consistent with the expectation for a circular orbit to within 3$\sigma$. The posterior probability distribution for HAT-P-38b's eclipse phase is bimodal in the version of the fits where the per-point errors are left as free parameters, with one peak within 2$\sigma$ of the predicted phase for a circular orbit and one peak corresponding to a secondary eclipse occurring $\sim$30 minutes later than expected. The peak corresponding to a circular orbit is the taller of the two peaks in the initial fit, and when we inflate the per-point errors for the visits with significant correlated noise this secondary peak is further suppressed, indicating that it is likely an artifact of the correlated noise. We therefore present the solution corresponding to the higher peak centered near a phase of 0.5 in Table~\ref{table:bestfit}. 

Our best fit solution for WASP-7b favors an eclipse that occurs 78.8$^{+5.0}_{-4.2}$ minutes early, corresponding to an $e$cos ($\omega$) of -0.0173$^{+0.0011}_{-0.0009}$ where $e$ is the orbital eccentricity and $\omega$ is the longitude of periastron. This time offset cannot be due to uncertainties in the planet's ephemeris, as the predicted time of secondary eclipse for a circular orbit has an uncertainty of less than 1 minute (Table ~\ref{table:systems1}). It is somewhat surprising that this planet would have an eccentric orbit, as the tidal circularization timescale for this system is predicted to be short ($\tau$$_{circ}$$<$650 Myr estimated using Eq. 2 from \cite{Bodenheimer2001} and a tidal quality factor $Q$=10$^{6}$). This is significantly shorter than the system's 2.4 Gyr estimated age, and WASP-7b does not appear to have an exterior companion capable of maintaining a non-zero eccentricity in the face of ongoing circularization.

We can use the difference in the brightness temperatures between the two bandpasses to ascertain whether we detect changes in a planet's emission spectrum due to the presence of spectral features. For HAT-P-38b and WASP-72b, which have non-detections in one of the two bandpasses, we take a conservative 2$\sigma$ upper limit as the brightness temperature in the bandpass with the non-detection. We find that all 5 planets have spectral shapes that are consistent with those of a blackbody at the $3\sigma$ level.

\begin{figure*}[]
\begin{centering}
\includegraphics[width=1\textwidth]{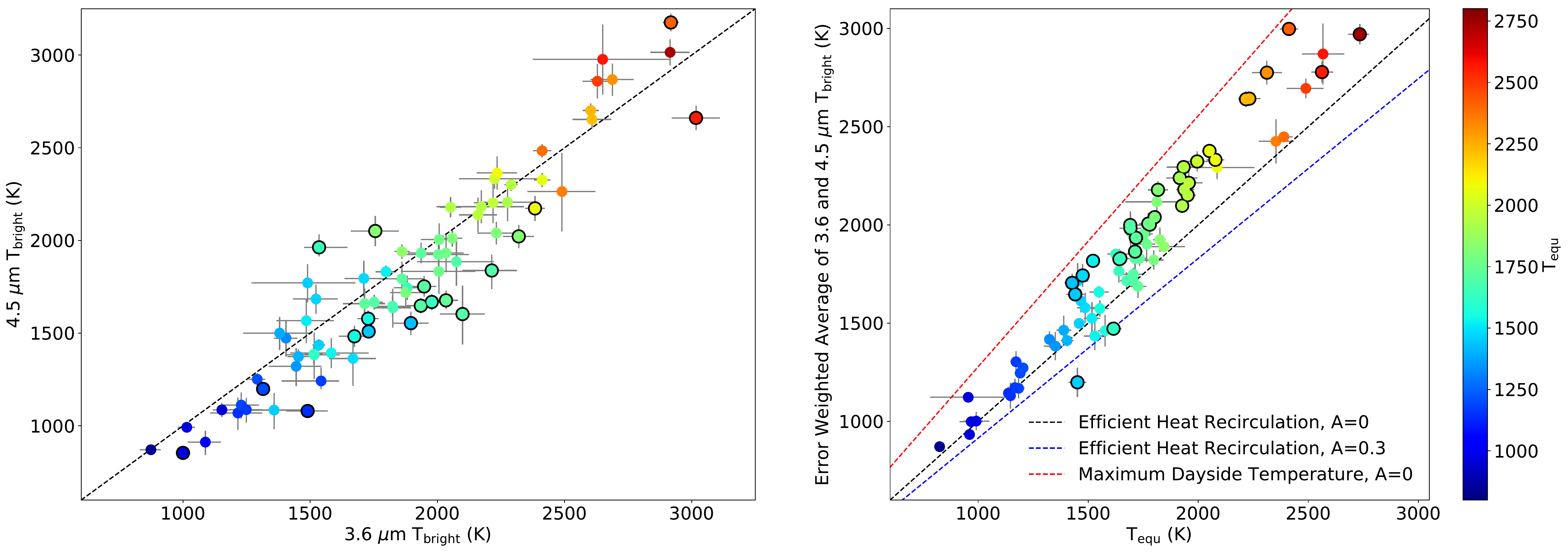}
\caption{Left: Measured 4.5 $\mu$m brightness temperature versus 3.6 $\mu$m brightness temperature for the sample of all planets with published detections in both bands as well as three new planets from this study. Planets with blackbody-like spectra will lie close to the black dashed line, which corresponds to a 1:1 brightness temperature ratio in the two bands. We highlight points that deviate from this line by greater than 3$\sigma$ (i.e., planets with non-blackbody emission) using black open circles. The color of the points indicates the predicted equilibrium temperature assuming efficient day-night circulation and zero albedo. Right: The error weighted average of the brightness temperatures measured in the 3.6 $\mu$m and 4.5 $\mu$m bandpasses versus the predicted equilibrium temperature. Planets that deviate by more than 3$\sigma$ from the dashed black line (the expected brightness temperature for efficient heat recirculation and zero albedo) are outlined in black.}
\label{fig:Temps}   
\end{centering}
\end{figure*}

\section{Discussion}\label{sec:discussion}

We add three of our new secondary eclipse observations to the sample of all \emph{Spitzer} secondary eclipse measurements from the literature. We include measurements of a subset of the 36 planets (27 of which were new) presented in \cite{Garhart2020} (those with detections in both bandpasses measured with a significance greater than 2.5$\sigma$; 31 planets). We also add 42 planets with published eclipses detected in both bandpasses at better than the 2.5$\sigma$ level (see Table~\ref{table:lit_planets} for the full list of included planets). With more than twice the number of planets, we are able to further investigate trends in the spectral shapes of short period gas giant planets with measured \emph{Spitzer} secondary eclipses. In order to be able to compare the thermal emission of these planets empirically, we calculate the brightness temperatures for each planet using the reported eclipse depths and planetary parameters and stellar parameters from \citealt{Southworth2011} (see Table~\ref{table:lit_planets} for more details).

\subsection{Trends as a Function of Incident Flux}\label{sec:flux_trends}
In the left-hand panel of Figure~\ref{fig:Temps} we plot the 4.5 $\mu$m brightness temperature versus the 3.6 $\mu$m brightness temperature to see if the spectral slopes of these planets deviate from that of a blackbody in a way that correlates with the incident flux. \cite{Garhart2020} found evidence for such a correlation in their study, but concluded that the trends they saw were not well-matched by commonly used model atmosphere grids. \cite{Baxter2020} used an expanded \emph{Spitzer} data set to reproduce the trend found by \cite{Garhart2020} and concluded that models including temperature inversions were better able to capture the qualitative shifts in spectral shape as a function of temperature. As shown in the left-hand panel of Figure~\ref{fig:Temps}, we also find that the most highly irradiated planets tend to lie above the line (meaning their 4.5 $\mu$m brightness temperature is higher than their 3.6 $\mu$m brightness temperature). This is consistent with the predictions of models presented in \cite{Lothringer2018}, who showed that planets in this temperature regime should have thermal inversions and additional opacity sources, such as $H^{-}$, that are not present in cooler atmospheres. At lower temperatures most of the planets in our sample tend to lie below this line, indicating that their brightness temperatures are relatively high at 3.6 $\mu$m and low at 4.5 $\mu$m. This is broadly consistent with the predictions of standard atmosphere models (\citealt{Burrows1997}, \citeyear{Burrows2006}; \citealt{Fortney2005}, \citealt{Fortney2008}), which suggest that the infrared spectra of these planets should be dominated by water and carbon monoxide absorption bands at these wavelengths. Because carbon monoxide overlaps significantly with the 4.5 $\mu$m \emph{Spitzer} band, these models predict that planets in this temperature range should have brightness temperatures that are lower at 4.5 $\mu$m than at 3.6 $\mu$m \citep{Garhart2020}. This picture changes for planets with temperatures less than 1000 K \citep{Wallack2019}, where methane is predicted to be the dominant carbon-bearing molecule. 

We can also use the difference between the band-averaged brightness temperatures and the expected equilibrium temperatures for each planet to investigate trends in circulation. As shown in the right-hand panel of Figure~\ref{fig:Temps}, the hottest planets in our sample appear to lie significantly above this line, indicating that they have less efficient day-night heat redistribution and low albedos. Indeed, none of these planets lie above the maximum dayside temperature line (calculated from \citealt{Pass2019}). As the incident flux decreases, planets move closer to the line corresponding to efficient day-night circulation, as predicted by general circulation models (\citealt{Perez-Becker2013}, \citealt{Komacek2016}). This is equivalent to the trend described in \cite{Schwartz2015}, where the hottest planets appear to have less efficient redistribution of heat. We quantify the significance of this trend using a Monte Carlo simulation where we generate one realization of the datapoints by sampling from the probability distributions for each point assuming all errors are reasonably well-approximated by Gaussian distributions, fitting a line to the resulting realization and repeating this 10$^{6}$ times. We find that the slope of the best-fit line differs from the equilibrium temperature expectation by greater than 6$\sigma$ and is qualitatively similar to the trend seen in \cite{Schwartz2015} despite the differences in the stellar models that we use. \cite{Schwartz2015} use a blackbody whereas we use a PHOENIX stellar model from \cite{Husser2013} integrated across the \emph{Spitzer} bandpass when determining the stellar flux.

We note that a recent study by \cite{Baxter2020} was unable to reproduce this trend in their compilation of \emph{Spitzer} secondary eclipse data when using a PHOENIX stellar model and integrating across the \emph{Spitzer} bandpass (although when using a blackbody as in \cite{Schwartz2015}, they were able to retrieve the trend). However, this study used the observed brightness temperature at 3.6 $\mu$m as opposed to the error weighted average of the brightness temperatures in the 3.6 and 4.5 $\mu$m bandpasses (C. Baxter, email communication). They argued that the presence of CO absorption in the 4.5 $\mu$m band makes the observed brightness temperature in this band more sensitive to potential changes in the dayside pressure-temperature profile than the 3.6 $\mu$m band. We evaluate the significance of the trend in our 3.6 $\mu$m data using the same Monte Carlo method as before. We find that using only the 3.6 $\mu$m data decreases the significance of the trend, but the best-fit line still deviates by $\sim$3$\sigma$ from the line defined by setting the brightness temperature equal to the equilibrium temperature. Although the slope of our best-fit line is consistent with the slope derived from \cite{Baxter2020} to 1$\sigma$, the error on our slope is a factor of $\sim$2 less, resulting in an increased significance.\footnote{We utilize the T$_\mathrm{equ}$ when calculating our slopes with effective temperature, but \cite{Baxter2020} utilize the T$_\mathrm{irradiation}$. This choice is inconsequential when evaluating the statistical significance of the slopes because the difference is just a factor of 1/$\sqrt{2}$, but we state this here for clarity.} The difference in uncertainties between our best-fit slope and that found in \cite{Baxter2020} is likely due to their use of orthogonal distance regression to fit the line and their inclusion of data with large errors (we chose to exclude planets with less than 2.5$\sigma$ detections in either the 3.6 or 4.5 $\mu$m bandpass).

Our study also obtains different dayside brightness temperatures than \cite{Baxter2020} for some individual planets. In some cases, this difference is due to the use of different values for the secondary eclipse depths (e.g. using \citealt{Knutson2012} instead of \citealt{Charbonneau2008} for the eclipse depth of HD 189733b). In other cases, it is due to the use of different values for the planet-star radius ratio and host star properties (effective temperature, metallicity, and gravity). In this study we use the most up-to-date values for the secondary eclipse depths, stellar parameters, and planetary parameters. Specifically, we source our stellar and planetary parameters from TEPCat, a database that seeks to compile the most recent and (where possible) homogeneously derived parameters for each exoplanetary system (\citealt{Southworth2008}, \citeyear{Southworth2009}, \citeyear{Southworth2010}, \citeyear{Southworth2011}, \citeyear{Southworth2012a}). Unlike \cite{Baxter2020}, we do not do an eccentricity cut on the planets that we include, but instead take the expected equilibrium temperature at the phase of secondary eclipse. We calculate our brightness temperatures using the same method as \citealt{Baxter2020} (i.e. using a PHOENIX stellar model from \cite{Husser2013} and integrating across each \emph{Spitzer} bandpass), and we obtain equivalent brightness temperatures when using the same input values.

We conclude that our data provide convincing evidence for a temperature-dependent change in recirculation efficiency. Although this trend might alternatively be interpreted as an increase in the dayside albedos of cooler planets, optical secondary eclipse measurements from \emph{Kepler} suggest that most planets in this temperature regime have relatively low albedos \citep{Heng2013}. There are two planets (WASP-94Ab and WASP-131b) that lie more than $3\sigma$ below the zero albedo efficient circulation line, suggesting that they may have appreciably higher albedos than the other planets in this sample. Both WASP-94Ab and WASP-131b lie in the same temperature regime as Kepler-7b (\citealt{Demory2011}, \citeyear{Demory2013}) and HATS-11b \citep{Niraula2018}, both of which have estimated geometric albedos of approximately 0.3 in the optical \emph{Kepler} bandpass (versus $<0.1$ for most gas giant planets in this temperature regime), likely due to the presence of high altitude silicate clouds \citep{Demory2013}. If WASP-94Ab and WASP-131b also have optical albedos close to 0.3, this would be sufficient to explain their lower than expected brightness temperatures. 

\begin{figure*}[htb!]
\includegraphics[width=1\textwidth]{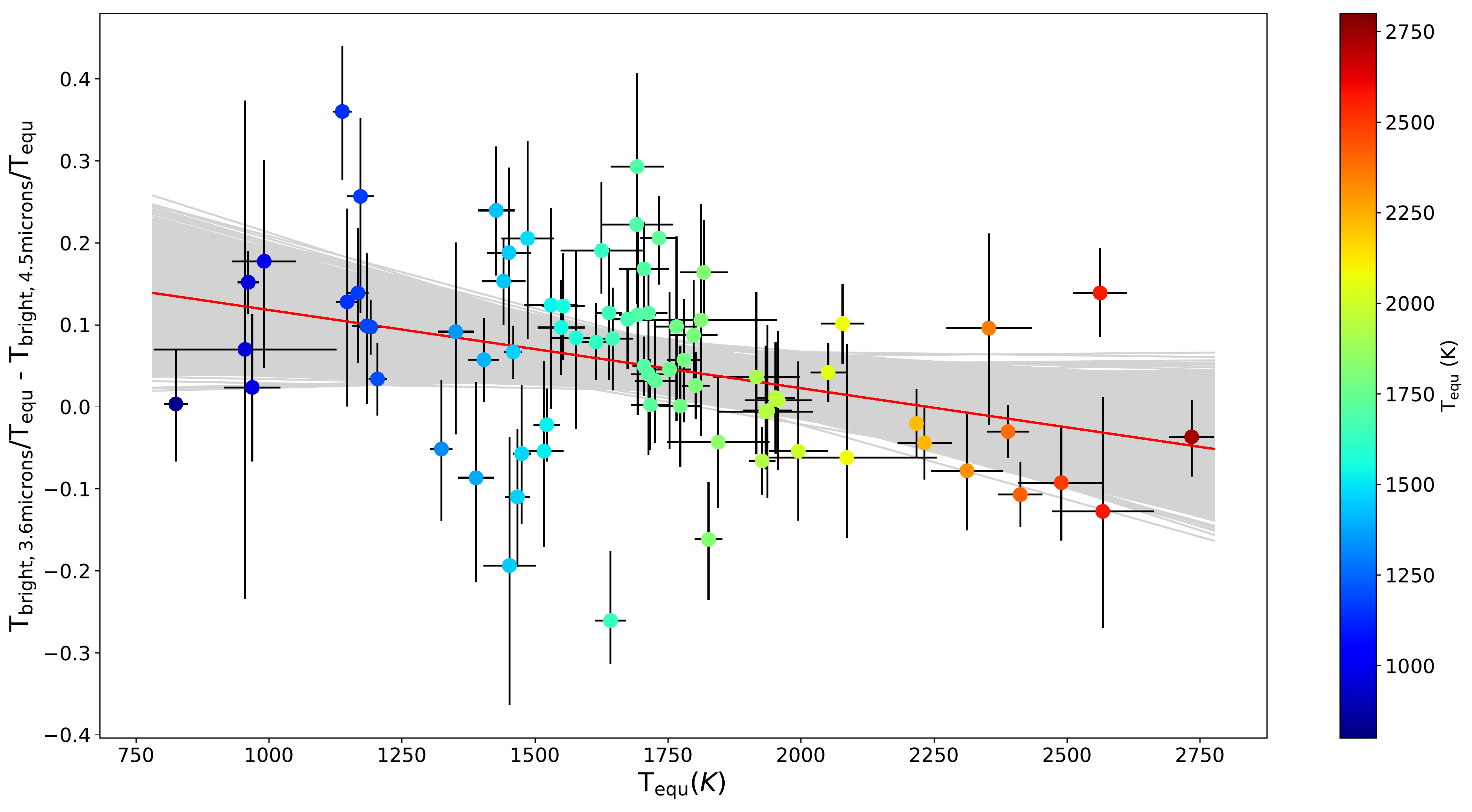}
\caption{Change in 3.6$-$4.5 $\mu$m slope as a function of equilibrium temperature, where slopes have been normalized by the equilibrium temperature to keep the scale consistent across the full temperature range. The best fit linear trend is overplotted as a red line with a random sampling of the fit lines from the Monte Carlo shown as gray lines (see \S\ref{sec:flux_trends} for a description of the Monte Carlo simulations).}
\label{fig:Tequ}   
\end{figure*} 

In order to better visualize the changes in the $3.6-4.5$ $\mu$m spectral slope as a function of incident flux, we calculate the difference in brightness temperatures between the two bands and divide by the equilibrium temperature in order to keep the scale of this slope constant across the full temperature range. We plot the resulting scaled slope as a function of the predicted equilibrium temperature assuming zero albedo and efficient circulation in Figure~\ref{fig:Tequ}. Unsurprisingly, the incident flux appears to be the primary driver of the observed spectral slope. We use the same Monte Carlo method as before to determine the significance of the trend in Figure~\ref{fig:Tequ}.We find that negatively sloped lines are preferred at the 3.2$\sigma$ level, suggesting that cooler planets tend to be brighter at 3.6 $\mu$m and dimmer at 4.5 $\mu$m than their more highly irradiated counterparts. 

\begin{figure*}[htb!]
\includegraphics[width=1\textwidth]{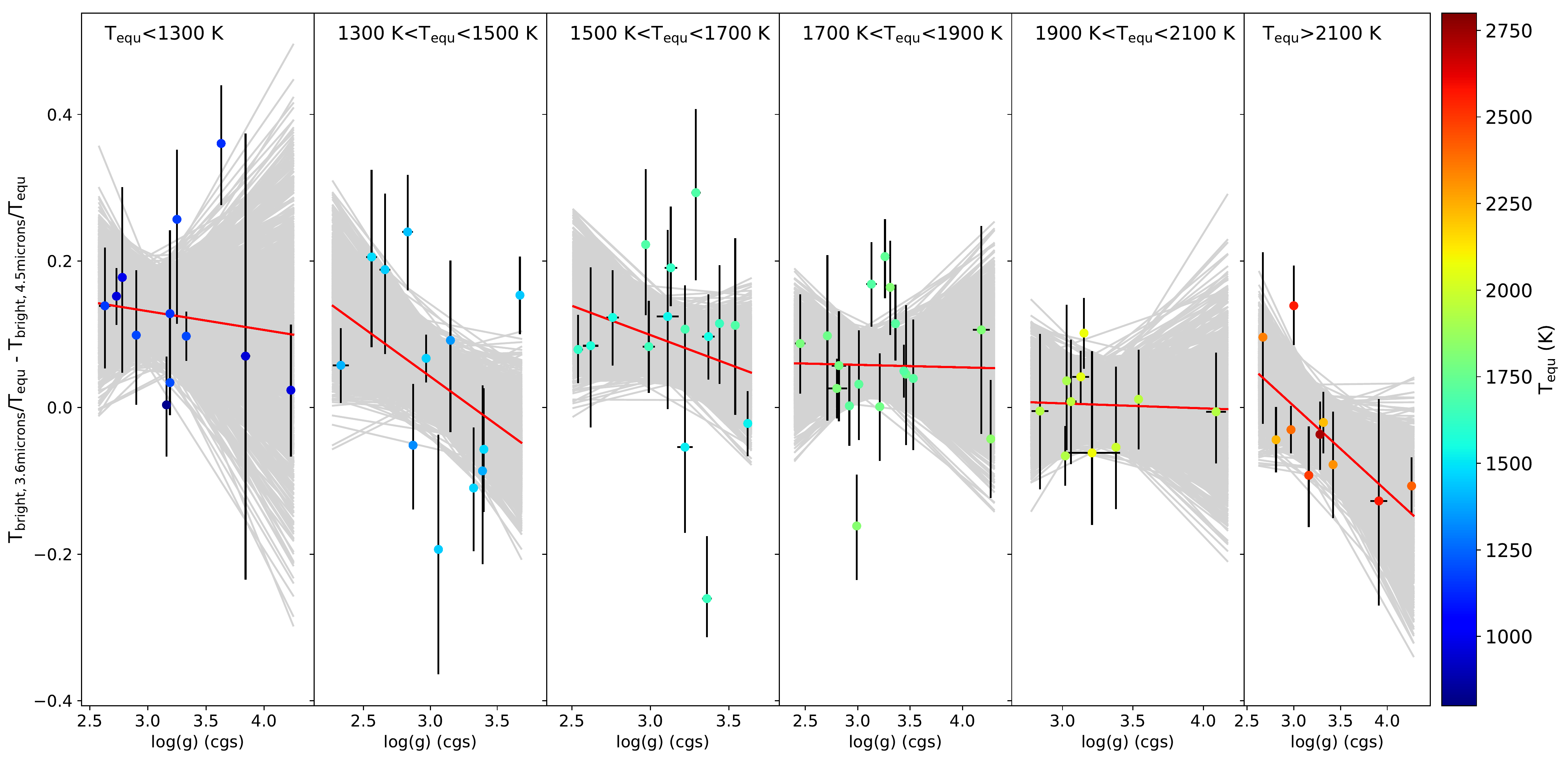}
\caption{The difference in the two bandpasses, each normalized by equilibrium temperature, versus the planet gravity, divided into temperature bins. The red lines are the best fit from the Monte Carlo simulations and a random sampling of the fit lines from the Monte Carlo are shown as gray lines (see \S\ref{sec:g_star_trends} for a description of the Monte Carlo simulations).}
\label{fig:gravity}   
\end{figure*} 

\subsection{Trends as a Function of Surface Gravity and Host Star Metallicity}\label{sec:g_star_trends}

We next consider whether or not there are additional parameters beyond incident flux that correlate with the observed spectral slopes of these planets. We divide our planet sample into bins according to their predicted equilibrium temperatures and investigate trends in spectral shape within each temperature bin. We first consider whether or not surface gravity can explain some of the observed scatter in the spectral shapes of these planets (see Figure~\ref{fig:gravity}). 

As before, we evaluate the significance of the trends in spectral slope as a function of surface gravity within each temperature bin using Monte Carlo simulations with 10$^{6}$ samples. We find that no temperature bin has a statistically significant slope (see Figure~\ref{fig:gravity}).

\begin{figure*}[htb!]
\includegraphics[width=1\textwidth]{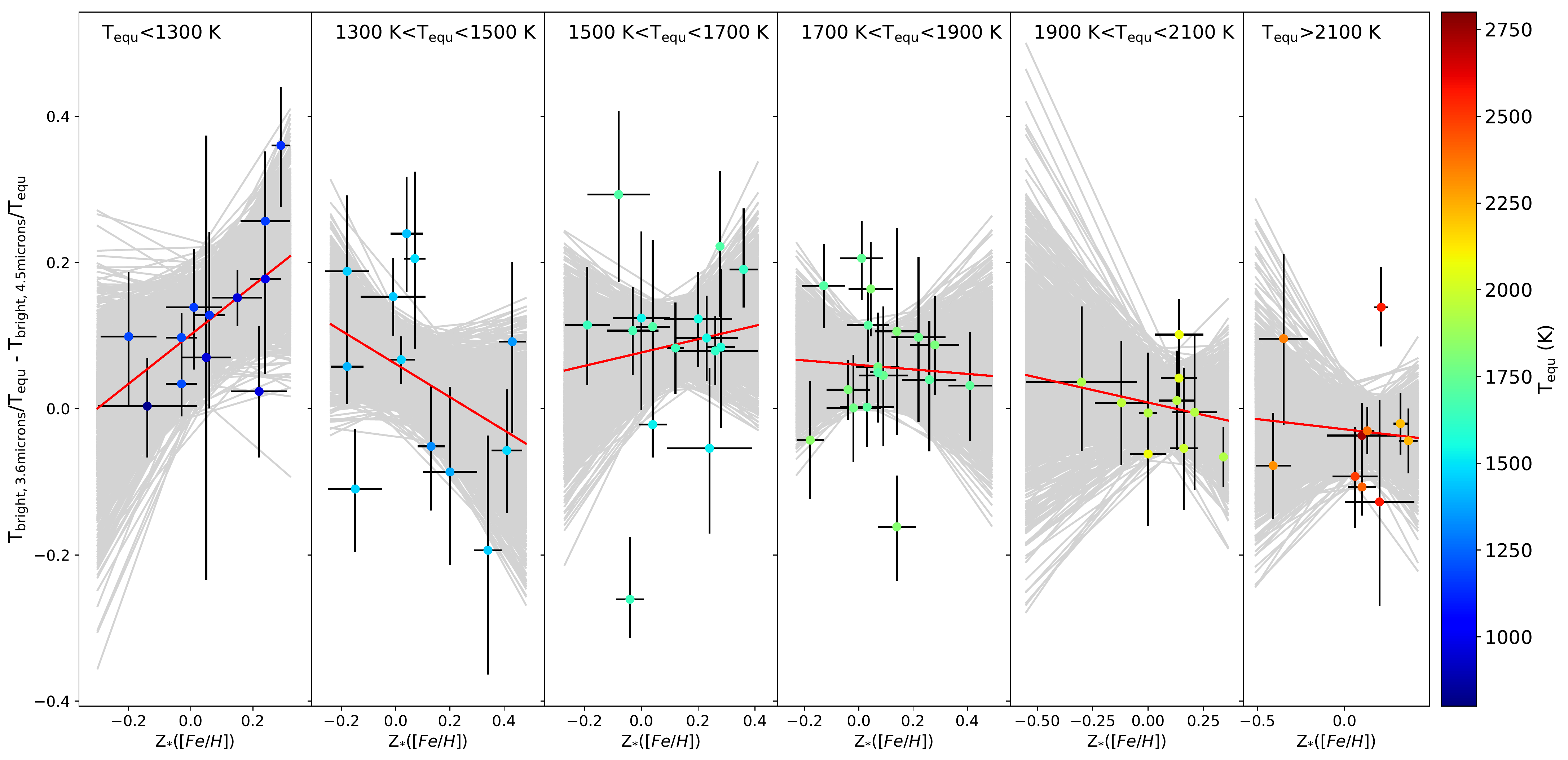}
\caption{The difference in the two bandpasses, each normalized by equilibrium temperature, versus the stellar metallicity, divided into temperature bins. The red lines are the best fit from the Monte Carlo simulations and a random sampling of the fit lines from the Monte Carlo are shown as gray lines (see \S\ref{sec:g_star_trends} for a description of the Monte Carlo simulations).}
\label{fig:cut_Z}   
\end{figure*} 

It is also possible that variations in atmospheric metallicity in this sample of planets might lead to variations in spectral slope. In \cite{Wallack2019}, we found a tentative ($\sim$1.9$\sigma$) correlation between the measured spectral slope of planets with equilibrium temperatures less than $\sim$1000 K and host stellar metallicity, which we use here as a proxy for planetary atmospheric metallicity. It is reasonable to assume that metal-rich stars should have correspondingly metal-rich disks, and therefore produce planets with correspondingly metal-enriched atmospheres. However, the range of metallicities spanned by the host stars in our planet sample only vary between $-0.41$ and $+0.43$, corresponding to a relatively small 6.9$\times$ change in bulk disk metallicity. In \cite{Wallack2019} we use a grid of atmosphere models to demonstrate that for cool ($\lesssim$ 1000 K) planets these \emph{Spitzer} observations would only be sensitive to changes in atmospheric metallicity that are one to two orders of magnitude larger. Indeed, there are other indications that this picture is not as straightforward as one might hope. For example, \cite{Teske2019} found that there are no statistically significant correlations between host star abundances and bulk planetary metallicity for planets less than 1000 K, further complicating attempts to link stellar metallicity to the resulting planetary atmospheric and bulk metallicities.

We use our new expanded sample to search for trends in stellar metallicity within each temperature bin (Figure~\ref{fig:cut_Z}). Extending the equilibrium temperature bin of the cooler planets to 1300 K, we also recover the relationship between spectral slope and stellar metallicity hinted at in \cite{Wallack2019}. In order to determine the significance of the trends in each panel in Figure~\ref{fig:cut_Z} we again use a Monte Carlo simulation where we sample from the probability distributions for each point. Interestingly, including these slightly warmer ($1000-1300$~K) planets in our lowest temperature bin 
does not greatly decrease the significance of the slope (which is still consistent with a flat line at the 1.8$\sigma$ level), despite using different sources for the stellar metallicities and including warmer planets. Although we would not necessarily expect planets warmer than 1000~K to have the same metallicity-dependent shift in the ratio of methane to carbon monoxide and carbon dioxide that we see in cooler atmospheres, our previous study (as well as \citealt{Drummond2018}) demonstrated that for planets with C/O ratios less than 1, there is still a detectable change in the $3.6-4.5$ $\mu$m spectral slope for these slightly warmer planets (see \citealt{Wallack2019} Figure 6).

 In contrast to this result we find that negative slopes are preferred for planets in the 1300-1500 K bin, but our data are still consistent with a positive sloped or flat line at the 1.5$\sigma$ level. It would not be surprising if the effects of metallicity on the spectral slope varied between temperature bins. As discussed above, changes in metallicity can alter the relative abundances of key molecules including methane, water, carbon monoxide, and carbon dioxide. For planets with high altitude cloud layers, changes in the metal content of the atmosphere might change the number density, vertical distribution, and average sizes of cloud particles \citep{Morley2013}. In general, we would expect increased cloud opacity at low pressures to suppress the amplitude of molecular features in the planet's dayside emission spectrum, making it look more like a blackbody (e.g., \citealt{Morley2017}) at near-infrared wavelengths. Indeed, planets with equilibrium temperatures in the 1300-1500 K temperature range are expected to host high altitude silicate cloud layers, and we might therefore expect this temperature bin to be more sensitive to metallicity-dependent changes in cloud properties at these wavelengths. At mid-infrared wavelengths on the other hand, we might expect to directly see emission peaks due to the presence of silicate clouds \citep{Richardson2007}.
 
\section{Conclusions}\label{sec:conclusions}

We present new secondary eclipse depth measurements in the 3.6 and 4.5 $\mu$m \emph{Spitzer} bands for HAT-P-5b, HAT-P-38b, WASP-7b, WASP-72b, and WASP-127b. We find that HAT-P-5b, HAT-P-38b, WASP-72b, and WASP-127b have secondary eclipse times consistent with the prediction for a circular orbit, but WASP-7b appears to have a modest orbital eccentricity ($e$cos ($\omega$) =-0.0173$^{+0.0011}_{-0.0009}$).

We combine these new detections with a sample of 73 planets with published 3.6 and 4.5 $\mu$m eclipse depths in an effort to better understand trends in the spectral shapes of these planets as a function of irradiation, surface gravity, and host star metallicity. We find that incident flux is the single most important factor for determining the atmospheric chemistry and circulation patterns of short-period gas giant planets. Although we would also expect surface gravity and host star metallicity to play a secondary role, we do not find any compelling evidence for correlations with these parameters in the current sample of \emph{Spitzer} eclipses. Most planets in our sample with the same incident flux level have broadly similar spectral shapes, but our study also reveals a subset of planets that appear as outliers in these plots. For example, WASP-94Ab and WASP-131b appear to be cooler than expected and may have high reflective cloud layers in their dayside atmospheres. Such planets are particularly promising targets for the \emph{James Webb Space Telescope} (JWST), which is currently scheduled for launch in 2021. The increased aperture size and wavelength coverage of \emph{JWST} will allow us to obtain invaluable new insights into the atmospheric compositions of gas giant planets.  

\acknowledgements
This work is based on observations made with the \textit{Spitzer Space Telescope}, which is operated by the Jet Propulsion Laboratory, California Institute of Technology under a contract with NASA. Support for this work was provided by NASA through an award issued by JPL/Caltech.  

\appendix
\section{}
We show the full list of planets (both from the literature and from this study) with detected eclipses (at the 2.5$\sigma$ level or better) in both the 3.6 $\mu$m and 4.5 $\mu$m \emph{Spitzer} bandpasses in the accompanying Table~\ref{table:lit_planets}. Bolded planets have new eclipses detected herein.

\startlongtable
\begin{deluxetable*}{lcccccc}
\tablecaption{Brightness Temperatures and System Parameters \label{table:lit_planets}}
\tablehead{
Planet   & $T_\textrm{equ}$ (K) &3.6 $T_\textrm{bright}$ (K) $\tablenotemark{a}$& 4.5 $T_\textrm{bright}$ (K) \tablenotemark{a}&log (gravity) (cgs) &[Fe/H]$_{*}$  & refs
}
\startdata
CoRoT-1b&1916$^{+81}_{-62}$&2276$^{+106}_{-109}$&2206$^{+101}_{-102}$&3.03$\pm$0.03&-0.30$\pm$0.25&1,2\\
CoRoT-2b&1522$\pm$25&1798$^{+40}_{-41}$&1831$^{+34}_{-35}$&3.62$\pm$0.02&0.04$\pm$0.05&1,2\\
HAT-P-1b&1324$\pm$21&1405$^{+45}_{-47}$&1473$^{+96}_{-101}$&2.87$\pm$0.01&0.13$\pm$0.05&1,3\\
HAT-P-2b\tablenotemark{b}&1812$\pm$143&2232$^{+74}_{-75}$&2040$\pm$61&4.18$\pm$0.08&0.14$\pm$0.08&1,4\\
HAT-P-3b&1172$\pm$26&1543$^{+71}_{-155}$&1242$^{+74}_{-44}$&3.25$\pm$0.03&0.24$\pm$0.08&1,5\\
HAT-P-4b&1691$^{+68}_{-32}$&2214$^{+99}_{-116}$&1838$^{+85}_{-101}$&2.97$^{+0.02}_{-0.04}$&0.277$\pm$0.007&1,5\\
\textbf{HAT-P-5b}&1517$\pm$37&1485$^{+109}_{-118}$&1567$^{+115}_{-121}$&3.22$\pm$0.05&0.24$\pm$0.15&1,6\\
HAT-P-6b&1705$\pm$47&1935$^{+55}_{-56}$&1648$\pm$41&3.13$\pm$0.05&-0.13$\pm$0.08&1,7\\
HAT-P-7b&2217$\pm$13&2608$^{+76}_{-77}$&2653$\pm$49&3.317$\pm$0.007&0.32$\pm$0.04&1,8\\
HAT-P-8b&1733$\pm$35&2034$^{+47}_{-68}$&1677$^{+53}_{-48}$&3.26$\pm$0.02&0.01$\pm$0.08&1,7\\
HAT-P-13b&1726$\pm$38&1714$^{+81}_{-85}$&1659$^{+82}_{-85}$&3.01$\pm$0.02&0.41$\pm$0.08&1,9\\
HAT-P-19b&991$^{+60}_{-58}$&1088$^{+61}_{-69}$&912$^{+62}_{-70}$&2.78$\pm$0.03&0.24$\pm$0.05&1,10\\
HAT-P-20b\tablenotemark{b}&969$\pm$53&1015$^{+32}_{-35}$&992$\pm$21&4.23$\pm$0.02&0.22$\pm$0.09&1,11\\
HAT-P-23b&1952$\pm$37&2160$^{+74}_{-75}$&2138$^{+93}_{-94}$&3.54$\pm$0.03&0.13$\pm$0.08&1,12\\
HAT-P-30b&1646$\pm$38&1882$^{+52}_{-53}$&1745$^{+66}_{-67}$&2.99$\pm$0.04&0.12$\pm$0.03&1,9\\
HAT-P-32b&1802$\pm$26&2059$^{+39}_{-40}$&2012$\pm$46&2.80$\pm$0.10&-0.04$\pm$0.08&1,13\\
HAT-P-33b&1780$\pm$32&2034$^{+69}_{-71}$&1932$^{+101}_{-104}$&2.82$\pm$0.06&0.07$\pm$0.08&1,9\\
HAT-P-40b&1766$^{+39}_{-44}$&2006$^{+143}_{-150}$&1833$^{+117}_{-121}$&2.71$\pm$0.03&0.22$\pm$0.10&1,9\\
HAT-P-41b&1937$^{+46}_{-38}$&2173$^{+168}_{-176}$&2182$^{+88}_{-90}$&2.84$\pm$0.06&0.21$\pm$0.10&1,9\\
HD149026b&1625$^{+77}_{-39}$&1978$\pm$37&1668$^{+44}_{-45}$&3.13$^{+0.03}_{-0.04}$&0.36$\pm$0.05&1,14\\
HD189733b&1191$\pm$25&1315$\pm$11&1199$\pm$9&3.33$\pm$0.02&-0.03$\pm$0.05&1,15\\
HD209458b\tablenotemark{c}&1459$\pm$17&1534$^{+24}_{-25}$&1436$^{+32}_{-33}$&2.968$\pm$0.004&0.02$\pm$0.05&1,16\\
KELT-2Ab&1713$^{+36}_{-29}$&1948$\pm$48&1752$^{+55}_{-56}$&3.36$\pm$0.04&0.034$\pm$0.078&1,9\\
KELT-3b&1817$^{+45}_{-46}$&2320$^{+60}_{-61}$&2022$^{+63}_{-64}$&3.31$\pm$0.04&0.044$^{+0.080}_{-0.082}$&1,9\\
KELT-7b&2051$\pm$33&2412$\pm$32&2326$\pm$38&3.13$\pm$0.06&0.139$^{+0.075}_{-0.081}$&1,9\\
Kepler-5b&1693$^{+35}_{-29}$&2075$^{+147}_{-154}$&1885$^{+125}_{-129}$&3.54$\pm$0.02&0.04$\pm$0.06&1,17\\
Kepler-6b&1452$^{+49}_{-26}$&1490$^{+188}_{-220}$&1771$^{+102}_{-105}$&3.06$^{+0.01}_{-0.02}$&0.34$\pm$0.05&1,17\\
Kepler-12b&1486$^{+49}_{-36}$&1668$^{+89}_{-94}$&1363$^{+135}_{-148}$&2.56$\pm$0.04&0.07$\pm$0.04&1,18\\
Kepler-13Ab&2567$\pm$96&2650$^{+265}_{-275}$&2977$^{+188}_{-190}$&3.91$\pm$0.09&0.20$\pm$0.20&1,19\\
Kepler-17b&1713$^{+33}_{-67}$&1861$^{+89}_{-93}$&1793$^{+93}_{-96}$&3.53$\pm$0.01&0.26$\pm$0.10&1,20\\
Qatar-1b&1389$^{+34}_{-33}$&1380$^{+127}_{-144}$&1500$^{+87}_{-91}$&3.39$\pm$0.01&0.20$\pm$0.10&1,9\\
TrES-1b&1147$\pm$21&1216$^{+96}_{-110}$&1069$^{+83}_{-92}$&3.19$\pm$0.03&0.06$\pm$0.05&1,21\\
TrES-2b&1467$\pm$23&1523$^{+86}_{-91}$&1684$^{+78}_{-80}$&3.323$\pm$0.006&-0.15$\pm$0.10&1,22\\
TrES-3b&1639$\pm$25&1825$^{+71}_{-73}$&1637$^{+103}_{-107}$&3.44$\pm$0.02&-0.19$\pm$0.08&1,23\\
TrES-4b&1798$\pm$45&1876$^{+61}_{-62}$&1719$^{+83}_{-85}$&2.45$\pm$0.05&0.28$\pm$0.09&1,24\\
WASP-1b&1826$^{+26}_{-32}$&1756$^{+92}_{-96}$&2051$^{+81}_{-82}$&2.99$^{+0.03}_{-0.04}$&0.14$\pm$0.07&1,25\\
WASP-3b&1995$^{+56}_{-48}$&2225$^{+192}_{-140}$&2333$\pm$54&3.38$\pm$0.03&0.161$\pm$0.063&1,26\\
WASP-4b&1674$\pm$29&1824$^{+70}_{-72}$&1645$^{+57}_{-58}$&3.221$\pm$0.009&-0.03$\pm$0.09&1,27\\
WASP-5b&1753$\pm$40&2004$^{+120}_{-125}$&1924$^{+94}_{-96}$&3.46$\pm$0.04&0.09$\pm$0.09&1,28\\
WASP-6b&1184$\pm$27&1229$^{+70}_{-77}$&1112$^{+68}_{-73}$&2.90$\pm$0.02&-0.20$\pm$0.09&1,10\\
\textbf{WASP-7b}&1530$\pm$50&1583$^{+147}_{-161}$&1393$^{+80}_{-82}$&3.11$\pm$0.07&0.00$\pm$0.10&1,6\\
WASP-8b\tablenotemark{b}&1138$\pm$17&1490$^{+79}_{-84}$&1080$^{+34}_{-36}$&3.63$\pm$0.02&0.29$\pm$0.03&1,29\\
WASP-10b\tablenotemark{b}&955$^{+172}_{-173}$&1153$^{+34}_{-36}$&1086$^{+38}_{-39}$&3.84$\pm$0.04&0.05$\pm$0.08&1,10\\
WASP-12b&2562$^{+51}_{-48}$&3017$^{+94}_{-95}$&2661$\pm$66&3.00$\pm$0.01&0.21$\pm$0.04&1,9\\
WASP-14b\tablenotemark{b}$^{,}$\tablenotemark{c}&1934$^{+89}_{-77}$&2290$\pm$26&2301$\pm$35&4.09$^{+0.08}_{-0.07}$&0.00$\pm$0.04&1,9\\
WASP-18b&2412$\pm$42&2918$\pm$32&3176$\pm$48&4.26$\pm$0.02&0.10$\pm$0.08&1,9\\
WASP-19b\tablenotemark{c}&2078$\pm$41&2384$\pm$39&2173$^{+67}_{-68}$&3.153$\pm$0.005&0.14$\pm$0.11&1,9\\
WASP-24b&1773$\pm$40&2007$^{+70}_{-71}$&2005$^{+88}_{-90}$&3.21$\pm$0.03&-0.02$\pm$0.10&1,30\\
WASP-33b&2734$^{+42}_{-53}$&2915$^{+77}_{-78}$&3015$\pm$70&3.28$\pm$0.04&0.10$\pm$0.20&1,14\\
WASP-39b&1167$\pm$20&1249$^{+61}_{-66}$&1087$^{+64}_{-69}$&2.63$\pm$0.05&0.01$\pm$0.09&1,10\\
WASP-43b&1441$\pm$41&1730$^{+24}_{-25}$&1509$^{+31}_{-32}$&3.67$\pm$0.01&-0.01$\pm$0.12&1,9\\
WASP-48b&1957$\pm$63&2219$^{+75}_{-76}$&2203$^{+108}_{-109}$&3.06$\pm$0.04&-0.12$\pm$0.12&1,12\\
WASP-62b&1427$\pm$35&1896$^{+70}_{-72}$&1554$^{+62}_{-64}$&2.83$\pm$0.04&0.04$\pm$0.06&1,9\\
WASP-63b&1577$^{+46}_{-32}$&1516$^{+101}_{-109}$&1383$^{+121}_{-131}$&2.62$\pm$0.05&0.28$\pm$0.05&1,9\\
WASP-64b&1692$\pm$50&2099$^{+87}_{-89}$&1603$^{+154}_{-164}$&3.29$\pm$0.03&-0.08$\pm$0.11&1,9\\
WASP-69b&961$\pm$20&1000$\pm$17&854$^{+18}_{-19}$&2.73$\pm$0.04&0.15$\pm$0.08&1,31\\
WASP-74b&1927$\pm$25&2052$\pm$41&2179$\pm$55&3.02$\pm$0.01&0.34$\pm$0.02&1,9\\
WASP-76b&2232$\pm$51&2603$\pm$32&2701$\pm$39&2.81$\pm$0.03&0.366$\pm$0.053&1,9\\
WASP-77Ab&1705$\pm$22&1752$\pm$35&1667$\pm$40&3.441$\pm$0.008&0.07$\pm$0.03&1,9\\
WASP-78b&2353$^{+81}_{-78}$&2490$^{+132}_{-136}$&2264$^{+208}_{-215}$&2.67$\pm$0.04&-0.35$\pm$0.14&1,9\\
WASP-79b&1717$^{+37}_{-34}$&1936$^{+51}_{-52}$&1932$\pm$56&2.92$\pm$0.04&0.03$\pm$0.10&1,9\\
WASP-80b&825$\pm$23&874$^{+39}_{-44}$&871$\pm$16&3.16$\pm$0.01&-0.14$\pm$0.16&1,32\\
WASP-87b&2312$\pm$68&2688$^{+84}_{-85}$&2868$^{+87}_{-88}$&3.42$\pm$0.03&-0.41$\pm$0.10&1,9\\
WASP-94Ab&1615$^{+36}_{-32}$&1514$^{+35}_{-36}$&1386$^{+50}_{-51}$&2.54$\pm$0.03&0.26$\pm$0.15&1,9\\
WASP-97b&1549$\pm$44&1728$^{+41}_{-42}$&1578$^{+44}_{-45}$&3.37$\pm$0.04&0.23$\pm$0.11&1,9\\
WASP-100b&2086$^{+169}_{-106}$&2235$^{+79}_{-81}$&2364$^{+89}_{-90}$&3.20$^{+0.20}_{-0.10}$&0.00$\pm$0.08&1,9\\
WASP-101b&1553$\pm$40&1674$^{+59}_{-61}$&1483$^{+57}_{-58}$&2.76$\pm$0.04&0.20$\pm$0.12&1,9\\
WASP-103b\tablenotemark{c}&2489$^{+81}_{-88}$&2629$^{+58}_{-59}$&2859$^{+92}_{-93}$&3.16$^{+0.02}_{-0.03}$&0.06$\pm$0.13&1,9\\
WASP-104b&1475$\pm$17&1711$^{+73}_{-76}$&1795$^{+95}_{-97}$&3.40$\pm$0.01&0.410$\pm$0.057&1,9\\
WASP-121b&2389$\pm$40&2412$\pm$36&2484$\pm$36&2.97$\pm$0.02&0.13$\pm$0.04&1,9\\
\textbf{WASP-127b}&1404$\pm$29&1454$^{+42}_{-43}$&1373$^{+40}_{-41}$&2.33$\pm$0.06&-0.18$\pm$0.06&1,6\\
WASP-131b&1451$\pm$41&1358$^{+109}_{-122}$&1085$^{+92}_{-103}$&2.66$\pm$0.04&-0.18$\pm$0.08&1,9\\
XO-1b&1204$\pm$17&1292$^{+32}_{-33}$&1251$^{+33}_{-34}$&3.19$\pm$0.01&-0.03$\pm$0.05&1,33\\
XO-2b&1351$^{+34}_{-53}$&1445$^{+98}_{-107}$&1321$^{+98}_{-106}$&3.15$\pm$0.03&0.43$\pm$0.05&1,34\\
XO-3b\tablenotemark{b}&1844$\pm$96&1861$\pm$30&1940$\pm$40 &4.27$\pm$0.03&-0.18$\pm$0.05&1,35\\
XO-4b&1642$^{+29}_{-25}$&1535$^{+112}_{-59}$&1963$^{+71}_{-50}$&3.36$\pm$0.03&-0.04$\pm$0.05&1,7\\
\enddata
\tablecomments{
\tablenotetext{a}{Planetary system parameter values used in computing brightness temperatures from \citet{Southworth2011}.}
\tablenotetext{b}{For this highly eccentric planet \citep{Bonomo2017a}, we report the temperature at the planet's separation at secondary eclipse.}
\tablenotetext{c}{We calculate the brightness temperature for this planet using the error-weighted average of the reported eclipses.}
}

\tablerefs{
 (1) \citet{Southworth2011}, (2) \citet{Deming2011}, (3) \citet{Todorov2010}, (4) \citet{Lewis2013}, (5) \citet{Todorov2013}, (6) this work, (7) \citet{Todorov2012}, (8) \citet{Wong2016}, (9) \citet{Garhart2020}, (10) \citet{Kammer2015}, (11) \citet{Deming2015}, (12) \citet{ORourke2014}, (13) \citet{Zhao2014}, (14) \citet{Zhang2018}, (15) \citet{Knutson2012}, (16) \citet{Evans2015}, (17) \citet{Desert2011b}, (18) \citet{Fortney2011}, (19) \citet{Shporer2014}, (20) \citet{Desert2011a}, (21) \citet{Cubillos2014}, (22) \citet{ODonovan2010}, (23) \citet{Fressin2010a}, (24) \citet{Knutson2009}, (25) \citet{Wheatley2010}, (26) \citet{Rostron2014}, (27) \citet{Beerer2011}, (28) \citet{Baskin2013}, (29) \citet{Cubillos2013}, (30) \citet{Smith2012}, (31) \citet{Wallack2019}, (32) \citet{Triaud2015}, (33) \citet{Machalek2008}, (34) \citet{Machalek2009}, (35) \citet{Machalek2010}
}
\end{deluxetable*}

\bibliography{references.bib}

\end{document}